\documentclass[12pt]{iopart}

\usepackage{graphicx,epsfig,latexsym,amssymb,color,dcolumn,bm}
\usepackage[bookmarksnumbered,plainpages]{hyperref}

\begin{document}

\title{Spherical-box approach for resonances in presence of Coulomb interaction}

\author{Shan-Gui Zhou$^{1,2,3}$, Jie Meng$^{4,5,1,2,3}$ and En-Guang Zhao$^{1,2,3,4}$}

\address{$^1$Key Laboratory of Frontiers in Theoretical Physics,
             Institute of Theoretical Physics, Chinese Academy of Sciences, Beijing 100190, China}
\address{$^2$Kavli Institute for Theoretical Physics China at the Chinese
             Academy of Sciences, Beijing 100190, China}
\address{$^3$Center of Theoretical Nuclear Physics, National Laboratory
             of Heavy Ion Accelerator, Lanzhou 730000, China}
\address{$^4$School of Physics, Peking University,
             Beijing 100871, China}
\address{$^5$Department of Physics, University of Stellenbosch, Stellenbosch,
             South Africa}

\ead{sgzhou@itp.ac.cn}

\begin{abstract}
The spherical-box approach is extended to calculate the resonance
parameters and the real part of the wave function for single
particle resonances in a potential containing the long-range Coulomb
interaction. A model potential is taken to demonstrate the ability
and accuracy of this approach. The calculated resonance parameters
are compared with available results from other methods. It is shown
that in the presence of the Coulomb interaction, the spherical-box
approach works well for not so broad resonances. In particular, for
very narrow resonances, the present method gives resonance
parameters in a very high precision.
\end{abstract}

\pacs{02.60.Lj, 31.10.+z, 32.70.Jz \\
(Some figures in this article are in colour only in the electronic
version)}

%02.60.Lj    Ordinary and partial differential equations; boundary
%            value problems
%31.15.-p    Calculations and mathematical techniques in atomic and
%            molecular physics (excluding electron correlation calculations)
%32.70.Jz    Line shapes, widths, and shifts

%\keywords{Resonance, real stabilization method, spherical-box,
%Coulomb interaction}

\date{\today}

\maketitle

%\tableofcontents

\section{INTRODUCTION}
\label{sec:intro}

The investigation of continuum and resonant states is an important
subject in quantum physics. For the theoretical determination of
resonant parameters (the energy and the width), in addition to many
approaches which are based on the connection of the unbound states
(including resonances and antibound or virtual states) and the poles
of the S matrix, several bound-state-like methods have been
developed, e.g., the complex scaling method (CSM)~\cite{Kato2006,
Kruppa1988, Reinhardt1982, Ho1983, Moiseyev1998, Gyarmati1986,
Kruppa1997, Moiseyev2005, Arai2006}, the complex absorbing potential
(CAP) method~\cite{Riss1993, Mueller2003}, the analytical
continuation in the coupling constant (ACCC)
approach~\cite{Kukulin1977, Kukulin1979, Kukulin1989, Tanaka1997,
Tanaka1999, Yang2001, Cattapan2000, Zhang2004, Guo2005}, and the
real stabilization method (RSM)~\cite{Hazi1970, Fels1971, Fels1972,
Maier1980, Taylor1976, Mandelshtam1993, Mandelshtam1994, Kruppa1999,
Adamson2008, Zhang2008}.

In the real stabilization method, the Schr\"{o}dinger equation (or
the Dirac equation in relativistic models) of the system in question
is solved in a basis~\cite{Hazi1970} or a box~\cite{Maier1980} of
finite size, thus a bound state problem is always imposed. The RSM
is based on the fact that the energy of a ``resonant'' state is
``stable'' against the change of the size of the basis or the box.
Many efforts have been made in order to calculate more efficiently
resonance parameters with the RSM~\cite{Maier1980, Taylor1976,
Mandelshtam1993, Mandelshtam1994, Kruppa1999, Adamson2008}. One of
the effective ways to do so is the so called ``spherical-box
approach''~\cite{Maier1980} in which the energy and width of a
resonance in finite-range potentials are determined from the
variation of the discrete positive energies with the radius of the
box. The spherical-box approach for short range potential has been
vigorously proved~\cite{Hagedorn2000}. In a recent
work~\cite{Zhang2008}, the single neutron resonances in atomic
nuclei were investigated by combining the RSM and the relativistic
mean field model~\cite{Vretenar2005, Meng2006}.

In many atomic, molecular and nuclear processes, the long-range
Coulomb interactions play important roles. Therefore it is necessary
to develop the spherical-box approach for resonances in the presence
of the long-range interaction and check its validity and accuracy.
In the present work, we extend the spherical-box approach in order
to include the influence of the Coulomb force.

The paper is organized as follows. In section~\ref{sec:formalism} we
give the formalism of the spherical-box approach for resonances in
the presence of the Coulomb interaction. The numerical details, the
results for a model potential and discussions are given in
section~\ref{sec:results}. Finally in section~\ref{sec:summary} we
summarize our work.

\section{Formalism}
\label{sec:formalism}

We deal with a central field problem. The radial Schr\"{o}dinger
equation reads (in atomic units)
\begin{equation}
 \left[ - \frac{1}{2}\frac{d^2}{dr^2} + \frac{l(l+1)}{2r^2} + V_0(r) + \frac{Z}{r}
 \right]
  \Psi_k(r)
 = E \Psi_k(r)
 \label{eq:schroedinger}
 ,
\end{equation}
where $V_0(r)$ is a finite-range potential and $E=k^2/2$. Under the
box boundary condition the continuum is discretized and one is left
with a bound state problem. When the size of the box ${\cal R}$ is
large enough, the energies of bound states, if there are any, do not
change with ${\cal R}$. In the continuum region, there are some
states stable against the change of the size of the box, i.e., the
energy of each of such states is almost constant with the variation
of ${\cal R}$; such stable states correspond to resonances. It
should be noted that the spherical-box approach is essentially
different from the ``variable phase approach'' where the potential
in question is amputated of its part extending beyond a distance
$\bar r$ and one needs to solve a series of scattering problems for
different $\bar r$~\cite{Calogero1967}.

In the spherical-box approach, the resonance energy is determined by
the stability condition,
\begin{equation}
 \left. \frac{\partial^2 E}{\partial {\cal R}^2} \right|_{E=E_\gamma} = 0
 \label{eq:energy}
 ,
\end{equation}
the corresponding box size is labeled as $\bar {\cal R}$, i.e.,
$E_\gamma=E(\bar {\cal R})$.

The width can be evaluated from the stability behavior of the
positive energy state against the change of the box size around
$\bar {\cal R}$~\cite{Maier1980}. When $r$ is large enough, the
finite-range potential $V_0(r)$ vanishes and $\Psi_k(r)$ satisfies,
\begin{equation}
 \Psi_k(r) \propto \frac{1}{k r}
  \sin(k r - \frac{l\pi}{2} - \gamma\ln(2kr) + \eta_l)
 ,
 \label{eq:scattering}
\end{equation}
where $\eta_l$ is the phase shift. $- \gamma\ln(2kr)$ comes from the
long range Coulomb interaction with $\gamma = Z/k$. The box boundary
condition $\Psi_k({\cal R}) = 0$ gives
\begin{equation}
 k {\cal R} - \frac{l\pi}{2}  - \gamma\ln(2k {\cal R}) + \eta_l = n\pi .
 \label{eq:node}
\end{equation}
Thus the derivative of the phase shift with respect to the box size
reads,
\begin{eqnarray}
 \frac{d\eta_l}{d{\cal R}}
 & = &
 -\left( 1 - \frac{\gamma}{ \sqrt{2E} {\cal R} }\right)
 \left( \sqrt{2E} + \frac{1}{\sqrt{2E}} \frac{dE}{d{\cal R}} {\cal R} \right)
 + \ln\left(2k{\cal R}\right) \frac{d\gamma}{d{\cal R}}
 .
 \label{eq:derivative1}
\end{eqnarray}
Around an isolated resonance, the energy $E$ and the phase shift
$\eta_l(E)$ satisfy the following relation,
\begin{equation}
 \label{eq:E-eta}
 \eta_l(E) = \eta_{l,\mathrm{pot}}(E)
         + \tan^{-1}\left( \frac{\Gamma/2}{E-E_\gamma} \right)
 .
 \label{eq:phase_shift}
\end{equation}
Under the assumption that the phase shift from the potential
scattering $\eta_{l,\mathrm{pot}}(E)$ varies slowly with respect to
the box size, i.e, $\partial\eta_{l,\mathrm{pot}}/\partial {\cal R}
\sim 0$, one derives the formula for the width,
%\begin{widetext}
\begin{equation}
 \frac{\Gamma}{2} =
 \frac{-\sqrt{2E_\gamma}}
      {\left( 1 - \frac{Z}{ \sqrt{2E_\gamma} \bar{\cal R} }\right)
       \left( \bar{\cal R}
            + 2E_\gamma \left[ \left. \frac{dE}{d{\cal R}} \right|_{{\cal R} = \bar{\cal R}}
                        \right]^{-1}
       \right)
       + \frac{Z}{2E_\gamma} \ln(\sqrt{8E_\gamma}{\cal R})
      }
 \label{eq:Gamma}
 \ .
\end{equation}
%\end{widetext}
Note that the above formula goes back to equation (8) in
\cite{Maier1980} when the Coulomb interaction is absent, i.e.,
$Z=0$.

\section{A model problem}
\label{sec:results}

In order to check the ability and the accuracy of the spherical-box
approach for resonances in potentials containing a long-range
interaction, we solve a model problem and compare our results with
those predicted by other methods. We choose the following potential,
\begin{equation}
 V(r) = V_0 r^2 e^{-r} + \frac{Z}{r}
 \label{eq:potl}
 ,
\end{equation}
which has been extensively investigated for $Z=0$ and/or
$Z=-1$~\cite{Isaacson1978, Maier1980, Mandelshtam1993, Yamani1995,
Sofianos1997}. We take $V_0=7.5$ in the present work.

\subsection{Numerical procedure}
\label{subsec:num}

The Schr\"{o}dinger equation (\ref{eq:schroedinger}) with the
potential (\ref{eq:potl}) is solved under the box boundary condition
by using the shooting method~\cite[Chapter 18]{Press2007} with the
forth order Runge-Kutta algorithm. For $Z=0$, the energy calculated
from the shooting method converges with the step size decreasing and
the relative error reaches to within $10^{-8}$ with a step size
$\delta r=0.001$. For $Z=-1$, a step size $\delta r=0.0001$ gives
the relative accuracy $\le 10^{-8}$ in the energy. In the following
we shall present results for the potential (\ref{eq:potl}) with
$Z=-1$. The step size $\delta r=0.0001$ will be used in order to get
a high accuracy comparable to the results given in the literature.

\subsection{Resonance parameters}

\begin{figure}[htb]
\begin{indented}
\item[]\includegraphics[width=8.0cm]{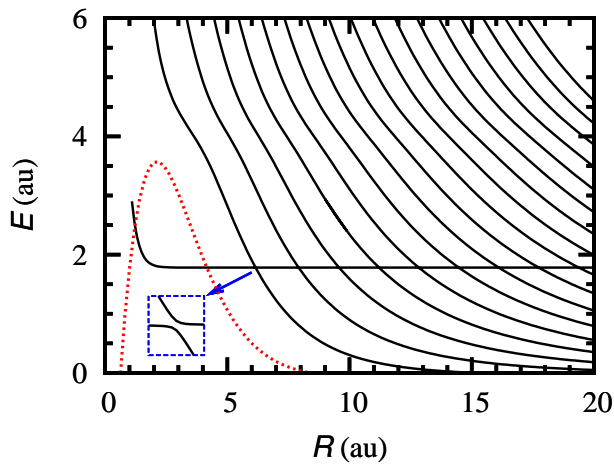}
\end{indented}
\caption{\label{fig:swave} The energies of s wave states calculated
in a spherical-box of different sizes. The inset shows the first
avoid crossing which is zoomed in. The red dotted curve represents
the potential (\ref{eq:potl}) around the barrier.}
\end{figure}

We first present results for s states. In figure~\ref{fig:swave} the
potential $V(r)$ given in (\ref{eq:potl}) and the energies of s
states versus the box size are presented. By examining the figure,
two resonances can be found with energies about 1.8 and 4.0
respectively. The former is very narrow as is indicated by the sharp
avoid crossings. The latter, however, is above the barrier and very
broad.

\begin{table}[htb]
\caption{\label{tab:1s} The energy and width of the first s wave
resonance evaluated at different avoid crossings from
(\protect\ref{eq:energy}) and (\protect\ref{eq:Gamma}). All
quantities are in atomic units.}
\begin{indented}
\item[]\begin{tabular}{r|l|l} \br
 $\bar{\cal R}\ \ $ & $\ \ E_\gamma$ & $\Gamma$ \\
\mr
 5.0163 & 1.780 531 212 & 6.302 9$\times10^{-5}$ \\
 7.0547 & 1.780 525 482 & 8.764 7$\times10^{-5}$ \\
 8.7509 & 1.780 524 837 & 9.344 1$\times10^{-5}$ \\
10.3918 & 1.780 524 706 & 9.507 7$\times10^{-5}$ \\
12.0226 & 1.780 524 661 & 9.553 6$\times10^{-5}$ \\
13.6542 & 1.780 524 606 & 9.566 2$\times10^{-5}$ \\
15.2878 & 1.780 524 620 & 9.569 6$\times10^{-5}$ \\
16.9243 & 1.780 524 629 & 9.570 7$\times10^{-5}$ \\
18.5633 & 1.780 524 635 & 9.571 1$\times10^{-5}$ \\
20.2045 & 1.780 524 634 & 9.571 3$\times10^{-5}$ \\
\mr
 %\footnote
 {\protect\cite{Sofianos1997}} & 1.780 524 536 & 9.571 9$\times 10^{-5}$ \\
 %\footnote
 {\protect\cite{Yamani1995}} & 1.780 5 & 9.58$\times 10^{-5}$ \\
\br
\end{tabular}
\end{indented}
\end{table}

The resonance parameters evaluated at different stable regions are
given in table~\ref{tab:1s} for the very narrow resonance. For
comparison, the results obtained from the complex methods are also
included~\cite{Yamani1995, Sofianos1997}. With $\bar{\cal R}$
increasing, the resonance energy converges to a stable value very
fast. For this narrow resonance, the spherical-box approach gives
very precise energy and width which are comparable to the exact
values given in~\cite{Sofianos1997}. One achieves a
seven-significant-digit accuracy for the energy at the third avoid
crossing with $\bar{\cal R} = 8.7509$ and a four-significant-digit
accuracy for the width at the ninth avoid crossing with $\bar{\cal
R} = 18.5633$. These high precisions are very encouraging. One can
then safely use this spherical-box approach in studying narrow
resonances with computational efforts much less than complex
calculations.

\begin{table}[htb]
\caption{\label{tab:2s} The energy and width of the second s wave
resonance evaluated at different avoid crossings from
(\protect\ref{eq:energy}) and (\protect\ref{eq:Gamma}). All
quantities are in atomic units.}
\begin{indented}
\item[]\begin{tabular}{r|l|l}
\br
 $\bar{\cal R}\ \ $ & $\ \ E_\gamma$ & $\Gamma$ \\
\mr
 3.4784 & 4.101 765 300 & 0.582 256 450 \\
 4.9943 & 4.053 255 228 & 0.804 538 485 \\
 6.2534 & 4.021 498 436 & 0.939 289 481 \\
 7.4361 & 3.995 348 518 & 1.024 872 144 \\
 8.5939 & 3.970 907 320 & 1.083 354 259 \\
 9.7483 & 3.945 799 270 & 1.101 418 144 \\
10.9117 & 3.917 591 013 & 1.144 053 591 \\
12.0986 & 3.881 431 163 & 1.198 720 256 \\
13.3808 & 3.807 212 130 & 1.329 057 297 \\
\mr
 %\footnote
 {\protect\cite{Sofianos1997}} & 4.101 494 946 & 1.157 254 428 \\
\br
\end{tabular}
\end{indented}
\end{table}

The second s wave resonance is quite broad. In table~\ref{tab:2s} we
list the resonance parameters obtained for this resonant state at
different stable points $\bar{\cal R}$. When ${\cal R} > 13.3808$,
one can not find a stable region satisfying the condition
$\partial^2 E /
\partial {\cal R}^2$ = 0 in the present numerical accuracy.
At the last observed avoid crossing, the resonance energy deviates
from the exact value by 8\% and the width by
15\%~\cite{Sofianos1997}. This may be used as a guide to estimate
the accuracy of the spherical-box approach for broad resonances.

A third s wave resonance was predicted to be lying at 4.66 by the
complex method~\cite{Sofianos1997}. It is very broad with a width
5.34. There is no hint for this resonance from
figure~\ref{fig:swave}. This implies that such a broad resonance is
beyond the ability of the spherical-box approach. We note that for
very broad resonances, the assumption that the phase shift from the
potential scattering $\eta_{l,\mathrm{pot}}(E)$ in
(\ref{eq:phase_shift}) varies slowly with respect to the box size
may not hold. This could be one of the reasons why the present
method does not work well for broad resonances.

%\subsubsection{p wave resonance}

\begin{figure}[htb]
\begin{indented}
\item[]\includegraphics[width=8.0cm]{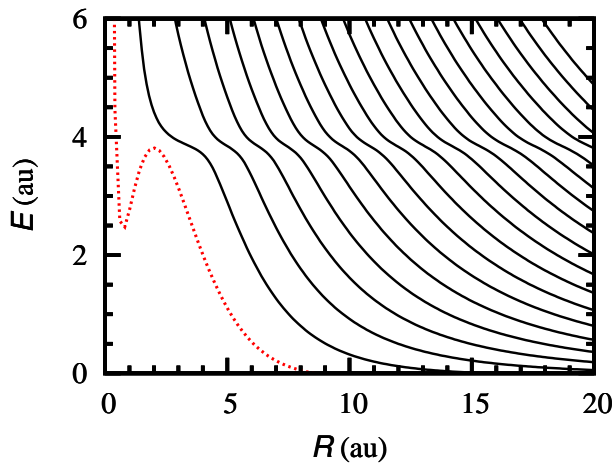}
\end{indented}
\caption{\label{fig:pwave} The energies of p wave states calculated
in a spherical-box of different sizes. The red dotted curve
represents the potential (\ref{eq:potl}) together with the
centrifugal potential.}
\end{figure}

\begin{table}[htb]
\caption{\label{tab:1p} The energy and width of the p wave resonance
evaluated at different avoid crossings from
(\protect\ref{eq:energy}) and (\protect\ref{eq:Gamma}). All
quantities are in atomic units.}
\begin{indented}
\item[]\begin{tabular}{r|l|l}
\br
 $\bar{\cal R}\ \ $ & $\ \ E_\gamma$ & $\Gamma$ \\
\mr
 3.3342 & 3.8665 & 0.1401 \\
 4.9740 & 3.8501 & 0.2101 \\
 6.2430 & 3.8471 & 0.2388 \\
 7.4205 & 3.8457 & 0.2519 \\
 8.5641 & 3.8450 & 0.2581 \\
 9.6945 & 3.8445 & 0.2610 \\
10.8201 & 3.8442 & 0.2608 \\
11.9441 & 3.8440 & 0.2616 \\
13.0681 & 3.8437 & 0.2620 \\
14.1925 & 3.8431 & 0.2623 \\
15.3174 & 3.8433 & 0.2625 \\
16.4430 & 3.8431 & 0.2626 \\
17.5691 & 3.8429 & 0.2628 \\
18.6958 & 3.8427 & 0.2629 \\
19.8229 & 3.8425 & 0.2631 \\
\br
\end{tabular}
\end{indented}
\end{table}

The energies of p states are plotted as a function of the box size
${\cal R}$ in figure~\ref{fig:pwave}. The effective potential
$V_\mathrm{eff}(r) = V(r) + l(l+1)/2r^2$ with $V(r)$ given in
(\ref{eq:potl}) is also shown in the same figure. Only one resonance
is found which is close to the barrier and the energy is around 3.8.
This state is broader than the first but narrower than the second s
wave resonances. The parameters for this resonance are given in
table~\ref{tab:1p}. Both the energy and the width converge well with
$\bar{\cal R}$ increasing. At the sixth stable point with $\bar{\cal
R} = 9.6945$, the relative deviation of the energy from the
converged value (obtained from the last avoid crossing at around
19.8229) is less than 0.1\% and that of the width less than 1\%.

%\subsubsection{d wave resonance}

\begin{figure}[htb]
\begin{indented}
\item[]\includegraphics[width=8.0cm]{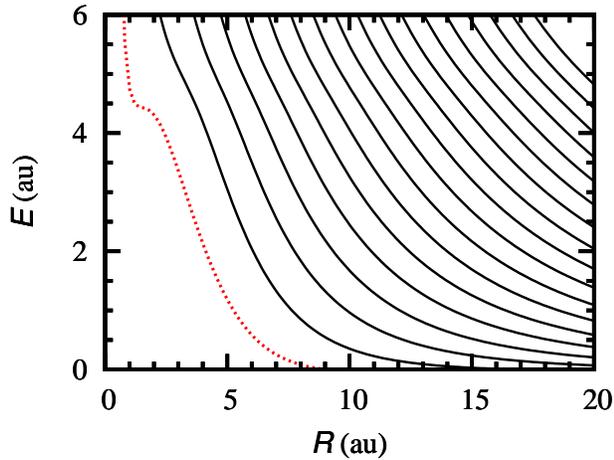}
\end{indented}
\caption{\label{fig:dwave} The energies of d wave states calculated
in a spherical-box of different sizes. The red dotted curve
represents the potential (\ref{eq:potl}) together with the
centrifugal potential.}
\end{figure}

Figure~\ref{fig:dwave} shows the $E\sim{\cal R}$ plot for the d
states. There is almost no pocket in the effective potential. But
one can still find slightly stable regions in the energy range
4.5$\sim$5.0 in the first several $E\sim{\cal R}$ curves. Indeed we
could find stable points fulfilling the condition $\partial^2 E /
\partial {\cal R}^2$ = 0 at the first six avoid crossings for the d
state. Thus the resonance parameters could be calculated and they
are listed in table~\ref{tab:1d}. At $\bar{\cal R} =9.3164$, the
relative deviation of the resonance energy is within 3\% and that of
the width is about 8\% compared with the values obtained at the
previous stable point $\bar{\cal R}= 8.1438$. Similar to the case
for the second s wave resonance, one can not find further stable
behavior in the region ${\cal R}> 9.3164$.

\begin{table}[htb]
\caption{\label{tab:1d} The energy and width of the d wave resonance
evaluated at different avoid crossings from
(\protect\ref{eq:energy}) and (\protect\ref{eq:Gamma}). All
quantities are in atomic units.}
\begin{indented}
\item[]\begin{tabular}{r|l|l}
\br
 $\bar{\cal R}\ \ $ & $\ \ E_\gamma$ & $\Gamma$ \\
\mr
 3.2624 & 4.8689 & 0.7135 \\
 4.6913 & 4.8063 & 0.9744 \\
 5.8924 & 4.7518 & 1.1486 \\
 7.0239 & 4.6992 & 1.2779 \\
 8.1438 & 4.6391 & 1.3920 \\
 9.3164 & 4.5409 & 1.5120 \\
\br
\end{tabular}
\end{indented}
\end{table}

%\subsubsection{Higher partial wave resonances}

No higher partial wave resonances are found in the present study.

\subsection{Wave functions of the resonances}

\begin{figure}[htb]
\begin{indented}
\item[]\includegraphics[width=8.0cm]{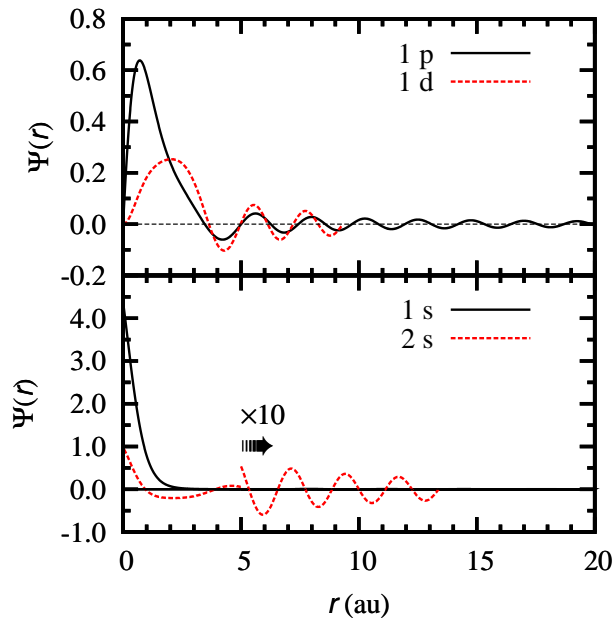}
\end{indented}
\caption{\label{fig:wf} The radial wave function $\Psi(r)$ of the
first and the second s wave resonances (lower panel) and of the p
and d wave resonances (upper panel). Note that when $r>5.0$, the
wave functions for the two s resonances are multiplied by 10.}
\end{figure}

Since the schr\"{o}dinger equation (\ref{eq:schroedinger}) is solved
within a spherical-box, we can only get the real part of the wave
function for each resonant states. The wave functions for the
resonances found in the present work are shown in
figure~\ref{fig:wf}. The behavior of the wave function is consistent
with the width of the resonance. For the first s wave resonance
which is quite narrow, the wave function is almost completely
localized inside the range of the finite range potential $V_0(r) =
7.5\ r^2 e^{-r}$. But for the broad s resonance, the radial wave
function inside the potential barrier is much depressed and it
oscillates very much in the asymptotic region. For the p wave
resonance, one still finds a localization feature in its radial wave
function, but not so prominent as the case of the first s resonant
state. The wave function of the d resonance is similar to that of
the second s state (note that the scales in the upper and lower
panels are different), except that the former vanishes at the
origin.

\begin{figure}[htb]
\begin{indented}
\item[]\includegraphics[width=8.0cm]{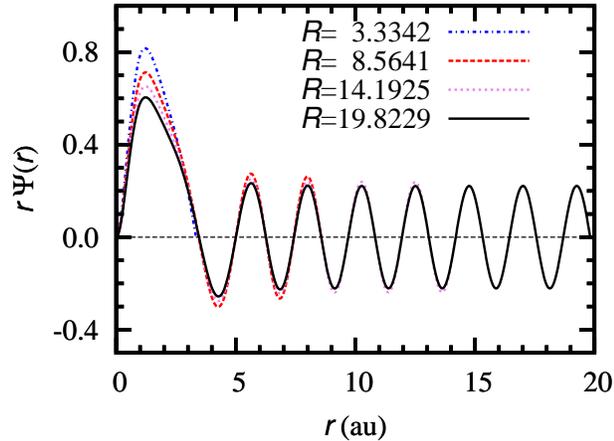}
\end{indented}
\caption{\label{fig:wf_p} The radial wave function of the p
resonance within a box of different size. Note that the wave
function is multiplied by the radial coordinate $r$ in order to see
clearly the asymptotic behavior. }
\end{figure}

The real stabilization method is based on the fact that a resonance
is more or less localized. Therefore the energy of a resonant state
is only weakly affected by the variation of the size of the
spherical-box. This has been shown in \cite{Maier1980, Zhang2008} as
well as in the present work. Next we show the convergence of the
real radial wave function with respect to the box size. The radial
wave function of the p wave resonance is shown in
figure~\ref{fig:wf_p}. In order to see clearly the asymptotic
behavior, the wave function $\Psi(r)$ is multiplied by the radial
coordinate $r$. With $\bar{\cal R}$ increasing, more nodes appear
and the wave function inside the potential barrier decreases only
slightly. This reveals that the wave function of a narrow resonance
is affected little by the box size ${\cal R}$ due to the
localization property.

\section{Summary}
\label{sec:summary}

The spherical box-approach, which is one of the effective
implementations of the real stabilization methods for single
particle resonances, is extended to the case in which a long-range
force such as the Coulomb interaction plays a role. The formalism is
presented and the numerical realization is fulfilled for this
approach.

We take a model potential (\ref{eq:potl}) as an example to
demonstrate the ability and the precision of this approach. It is
shown that in the presence of the Coulomb interaction, the
spherical-box approach still works well for narrow resonances. In
particular the present method can give resonance parameters in a
quite high precision for very narrow resonances. The energy and
width also converge reasonably fast with the box size increasing.
Within the spherical-box, the real wave functions of these
resonances are obtained. The localization behavior of the radial
wave function is consistent with the width of a resonance. The
convergence of the radial wave function with the increases of the
box size is studied, which shows that the wave function of a narrow
resonance is also ``stable'' against the changes of the box size.

\ack{We would like to thank Professor Shun-Jin Wang for helpful
discussions. This work was partly supported by the National Natural
Science Foundation (10705014, 10775004, and 10875157), the Major
State Basic Research Development Program of China (2007CB815000) and
the Knowledge Innovation Project of CAS (KJCX3-SYW-N02). The
computation of this work was supported by Supercomputing Center,
CNIC, CAS. }

\section*{References}

%\bibliographystyle{unsrt}
%\bibliographystyle{iopart-num}
%\bibliography{crsm}

\providecommand{\newblock}{}

\end{document}